\documentclass{INTERSPEECH2023}

\interspeechcameraready
\usepackage{enumitem}
\usepackage{multirow}
\usepackage{tablefootnote}
\usepackage{xcolor} 
\usepackage{eso-pic}
\usepackage{tikz}

\title{\textcolor{black}{A Compact End-to-End Model with Local and Global Context for Spoken Language Identification }}

\name{Fei Jia, Nithin Rao Koluguri, Jagadeesh Balam, Boris Ginsburg}
\address{NVIDIA Corporation, United States}
\email{fjia, nkoluguri, jbalam, bginsburg@nvidia.com}


\AddToShipoutPicture*{%
  \AtPageLowerLeft{%
    \begin{tikzpicture}[remember picture,overlay]
      \node[anchor=south west, text width=\paperwidth] at (1in,\dimexpr 1in - 2cm) {%
        To appear in \textit{Proc. INTERSPEECH 2023, August, 20-24, 2023, Dublin, Ireland}
      };
    \end{tikzpicture}%
  }%
}

\begin{document}

\maketitle

\begin{abstract}
We introduce \textcolor{black}{TitaNet-LID}, a compact end-to-end neural network for Spoken Language Identification (LID) that is based on the ContextNet architecture. 
TitaNet-LID employs 1D depth-wise separable convolutions and Squeeze-and-Excitation layers to effectively capture local and global context within an utterance.
Despite its small size, TitaNet-LID achieves performance similar to state-of-the-art models on the VoxLingua107 dataset while being 10 times smaller.
Furthermore, it can be easily adapted to new acoustic conditions and unseen languages through simple fine-tuning, achieving a state-of-the-art accuracy of 88.2\% on the FLEURS benchmark.
Our model is scalable and can achieve a better trade-off between accuracy and speed. TitaNet-LID performs well even on short utterances less than 5s in length, indicating its robustness to input length.
\end{abstract}

\section{\textcolor{black}{Introduction}}
\label{sec:intro}
Spoken Language Identification (LID)
is an important pre-processing step in various applications \textcolor{black}{such as Automatic Speech Recognition (ASR) and speech translation.} It requires robust performance while introducing minimal latency in the pipeline. 
\textcolor{black}{There are two important characteristics of LID models that have been rarely investigated. 
Firstly, LID models must be able to adapt easily to unseen languages and be robust to unseen domains.
Secondly, while the performance of the LID task is heavily influenced by speech duration and the amount of information present~\cite{rapid}, the model should still exhibit reliable performance on short speech segments.}

Early approaches to LID mainly relied on acoustic, phonetic, prosodic, morphologic, and semantic level representations~\cite{survey}.
However, recent advancements in deep neural networks have led to better performance \cite{Liu2022PHOLIDAU,mazzawi2019improving,shen2019Interactive,Shen2022TransducerbasedLE,bartley2023accidental,cai2019utterance,Cai2018ExploringTE} compared to traditional systems~\cite{langid_dnn}.
For instance, \cite{Gonzalez2014lstm} explores using Long Short Term Memory (LSTM) for LID task on NIST LRE 2009,
and demonstrates that it can exploit temporal dependencies effectively.
\cite{cnn_robust_interspeech} conducts experiments using bottleneck features from Convolutional Neural Network (CNN) on noisy RATS Radio traffic data~\cite{rat} and shows it has consistent improvement regardless of various acoustic conditions and durations.
\cite{miao19b_interspeech} proposes the addition of a time-frequency attention mechanism to Time Delay Neural Network architecture (TDNN) and CNN-LSTM-TDNN models to capture longer temporal dependencies, showing that frequency attention is more effective than time attention.
More recently, ECAPA-TDNN~\cite{desplanques2020ecapa} further improves TDNN performance on speaker tasks by introducing additional skip connections.
This architecture has been adopted to LID task, referred to as ECAPA-TDNN-LID~\cite{speechbrain} below, training on VoxLingua107~\cite{valk2021voxlingua107} and achieving 6.7\% error rate on the evaluation set. 

\textcolor{black}{Although CNN-based models have been effective in speech processing tasks, their limited kernel size often results in only capturing local context. This can be particularly detrimental for the LID task, where the amount of available context is critical. Incorporating global context models may address this limitation and improve the performance of CNN-based LID models. Among several proposed methods \cite{desplanques2020ecapa, gulati2020conformer, Radfar2022ConvRNNTCA, han2020contextnet}, ContextNet, proposed by \cite{han2020contextnet}, builds upon previous work \cite{kriman2020quartznet} and incorporates a Squeeze-and-Excitation (SE) layer \cite{hu2018squeeze} to compress a sequence of local feature vectors into a single global context vector. This global context vector is then broadcasted back to each local feature vector and combined via multiplications. In the ASR domain, \cite{han2020contextnet} demonstrated that incorporating global context improved accuracy. 
Another model, TitaNet~\cite{koluguri2022titanet}, utilizes the encoder of the ContextNet model as a top-level feature extractor, and applies an attentive pooling layer to capture utterance-level speaker representation, achieving outstanding performance in speaker recognition and diarization tasks. In this study, we aim to explore whether this encoder architecture can also enhance the performance for LID task.}
 
Another approach for LID is to first learn a good language representation vector, which can then be used for language category classification. To learn such a vector, one can use self-supervised learning (SSL).
Large scale speech understanding models like XLS-R (0.3B)~\cite{babu2021xls} and w2v-bert-51 (0.6B)~\cite{conneau2022xtreme} are able to achieve great results  by pre-training with more than 400k unlabeled data and then fine-tuning on LID datasets.
More recently, \cite{kukk2022accent} feeds the outputs of XLS-R (0.3B) model through an attentive pooling layer~\cite{zhu18_interspeech}, achieving state-of-the-art (SOTA) results of 4.7\% on VoxLingua107.
Though SSL-pretrained models show excellent performance, their enormous model size makes it challenging to deploy them in compute-constrained scenarios.

In this work we make the following contributions:
\begin{itemize}
   \item We propose \textit{TitaNet-LID}, a compact end-to-end neural network (NN) for LID, \textcolor{black}{which is based on ContextNet encoder, combining local features from 1D depth-wise separable convolutions and global context from Squeeze and Excitation (SE) layers~\cite{hu2018squeeze}. The decoder contains a statistic pooling layer followed by two linear layers. We open-source the model through NVIDIA NeMo~\cite{kuchaiev2019nemo}.\footnote{\url{https://github.com/NVIDIA/NeMo}}}
\item On VoxLingua107 dataset, we show that the model matches SOTA performance among other models that are trained from scratch. 
Its accuracy is similar to XLS-R model which uses SSL pretraining. Note that TitaNet-LID is 10X smaller than XLS-R. 
\item We examine TitaNet-LID's adaptation capability to unseen languages and new acoustic conditions. On the FLEURS dataset, TitaNet-LID outperforms the previous SOTA model \textcolor{black}{by 16.8\%} while using 20 times lesser number of parameters. 
\item We study the effect of segment length on accuracy  and demonstrate that model performs well even on short $\leq5s$ segments.
\item We show that the model can be easily scaled up or down for better accuracy or lower latency respectively. 
\end{itemize}

\begin{figure}
  \centering
  \centerline{\includegraphics[width=24em]{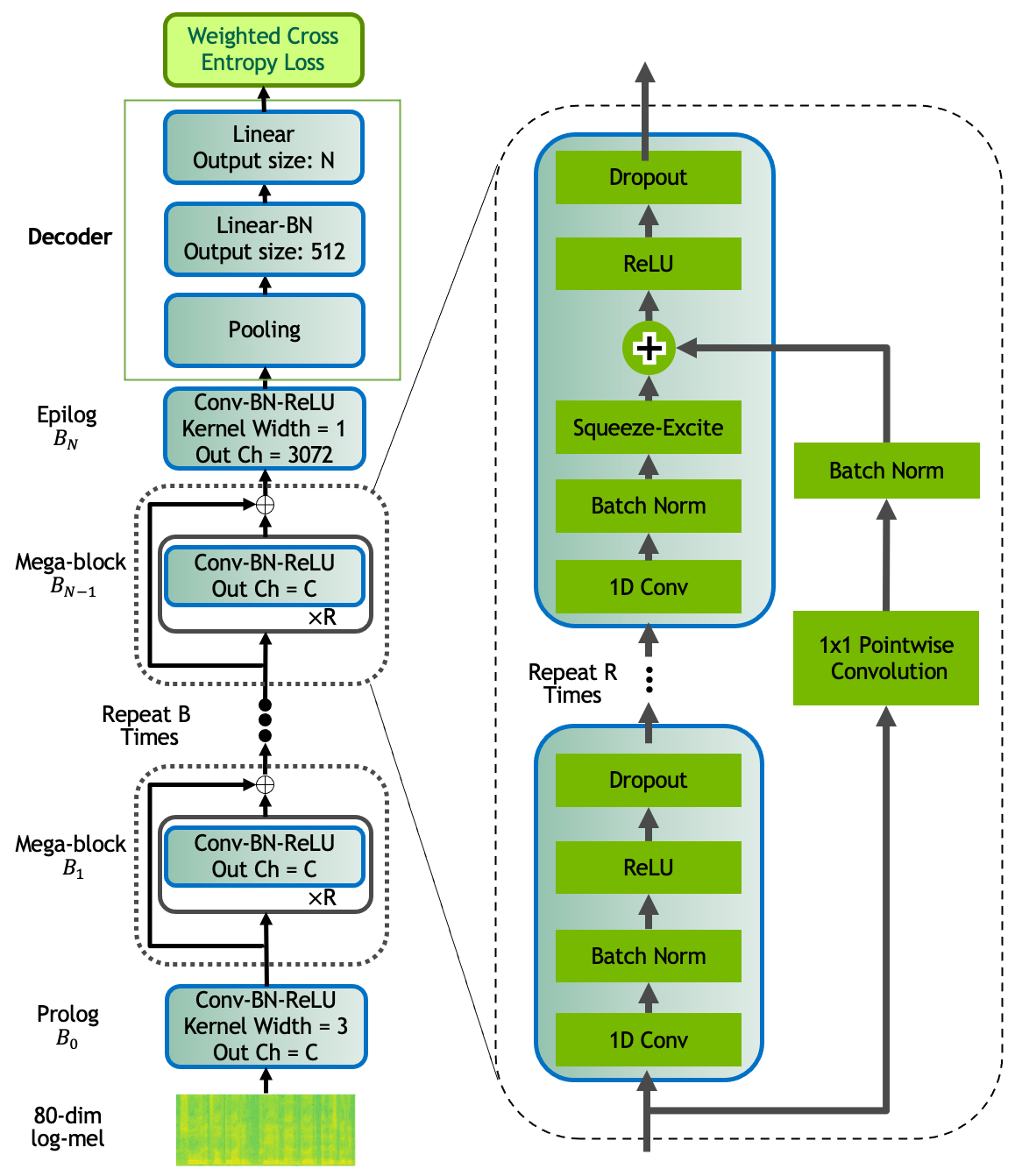}}
  \caption{\textcolor{black}{TitaNet-LID-$B$x$R$x$C$ Encoder is based on ContextNet architecture~\cite{han2020contextnet}, where $B$ is the number of blocks, $R$ is the number of repeated ``basic” blocks and $C$ is the number of filters in the convolution layers of each block. }}
  \label{model_arch}
\end{figure}


\section{MODEL ARCHITECTURE}
\label{sec:model}
\textcolor{black}{As illustrated in Figure~\ref{model_arch}, TitaNet-LID-$B$x$R$x$C$ consists of an encoder which is a 1D depth-wise channel separable convolutional model with a ContextNet-like~\cite{han2020contextnet,koluguri2022titanet} architecture and a decoder.
The model's encoder includes $B$ blocks, each block comprising $R$ repeated ``basic" blocks and $C$ filters in the convolution layers of each block.}

The TitaNet-LID encoder starts with a prologue block $B_0$, and followed by residual mega blocks $B_1...B_{N-1}$.
Each mega block consists of $R$ “basic” blocks and a squeeze-excite module in the end. A basic block is combined from 1D time-channel separable convolutional~\cite{kriman2020quartznet} (noted as 1D Conv in Figure~\ref{model_arch}) module with Kernel K, followed by BatchNorm, ReLU, and Dropout.
Each 1D Conv module consists of two parts: a depth-wise convolutional layer and a 1x1 pointwise convolutional layer. 
These basic blocks are repeated $R$ times and  connected residually with SE~\cite{hu2018squeeze} layers with global average pooling as in \cite{koluguri2022titanet, majumdar2021citrinet}. \textcolor{black}{The kernel K in repeated ``basic" blocks are 7,11,15.}
The encoder ends with an epilogue block $B_N$ and outputs intermediary audio features. 

The encoded audio features are passed to the decoder to perform classification. 
The decoder contains two parts: a statistics pooling layer (mean and std)~\cite{snyder2018x} to map variable length of input audio features to fixed-length feature representation, and two linear layers, one of output size 512 and another for a linear transformation from 512 to the final number of classes $N$.

\section{Experiment Setup and Results}
\label{sec:result}

\subsection{Train on VoxLingua107}
\label{ssec:voxlingua107}

\subsubsection{Dataset}
\label{sssec:voxlingua107_dataset}

VoxLingua107~\cite{valk2021voxlingua107} is a large dataset designed for LID task which contains diverse YouTube data for 107 languages. 
The dataset is quite noisy with samples containing synthetic speech, singing, yelling, etc.
The size of the official training set is 6628 hours, 62 hours per language on average, but it's highly imbalanced.

We split 10\% of data of each language in training set into validation set. 
The input utterances have been split into non-overlap 3-second segments resulting in 5.9M and 656K segments in train and validation set respectively.
The dataset comes with an official evaluation set of 1609 verified utterances covering 33 languages. 
The duration of evaluation samples varies from 1.7 seconds to 19.98 seconds. From 1609 samples, 253 are found to be less than 5 seconds.

\subsubsection{Training setup}
\label{sssec:voxlingua107_exp}
The audio segments are pre-processed into 80-dimensional log-melspectrograms calculated using 25ms window with stride of 10ms.
Due to the extreme imbalance of the number of samples for each language in training data, the TitaNet-LID model is trained with weighted cross entropy loss.
The weight of each class is calculated from data using
\begin{align}
    w_i = \frac{\sum_{n=0}^N{c_n}}{c_i},
\end{align}
where $N$ is number of classes and $c_{i}$ is number of samples of the given class $i$. The weights are then normalized to sum up to one. 
To value the minority class, macro accuracy which computes the accuracy for each class and then take the average are used during validation to select the best checkpoint for each run. 

All models were trained for 40 epochs on 4 nodes with 8 GPUs in  each node with a batch size of 128 per GPU.
We utilized the Adam optimizer and Cosine Annealing learning rate scheduler with a warm-up ratio of 10\%. A maximum learning rate of 0.001 and a minimum learning rate of 0.0001 were used. 
We apply noise augmentation, speed perturbation with 0.95x and 1.05x, and RIR impulse corpora~\cite{rir} perturbation as well as SpecAugment~\cite{park2019specaugment}.

\subsubsection{Results}
\label{sssec:voxlingua107_results}

We compare TitaNet-LID with two other models trained from scratch: Resnet34~\cite{valk2021voxlingua107} and ECAPA-TDNN-LID~\cite{speechbrain}. 
We also add a comparison with three pre-trained self-supervised learning (SSL) models finetuned on LID task: 
wav2vec2.0~\cite{babu2021xls}, XLS-R~\cite{babu2021xls} and XLS-R-attentive~\cite{kukk2022accent}. 
XLS-R~\cite{babu2021xls} is a cross-lingually wav2vec 2.0 model which has been pretrained on 436K hours of 128 languages data including VoxLingua107, and it has 300M parameters. It achieves 5.7\% error rate on the official evaluation set. 
XLS-R-attentive~\cite{kukk2022accent} extends XLS-R with an additional attentive pooling layer and reaches 4.7\% error rate. 
We have also experimented with attentive pooling layer in \cite{koluguri2022titanet} instead of statistic pooling layer but we don't observe performance gain. 

\begin{table}
\centering
\resizebox{\columnwidth}{!}{%
\begin{tabular}{lrrrr} \toprule
\multirow{2}{*}{Model}  & \multicolumn{1}{c}{params} & \multicolumn{3}{c}{ error rate (\%)} \\
                        &   \multicolumn{1}{c}{(M)}  & 0...5s & 5...20s  & avg  \\ \midrule
\textit{trained from scratch }  &                    &      &      &     \\
\quad Resnet34~\cite{valk2021voxlingua107}                      & 21    & 12.3          & 6.1         & 7.1      \\
\quad ECAPA-TDNN-LID~\cite{speechbrain}\tablefootnote{The number of parameters and error rate are obtained with checkpoint https://huggingface.co/speechbrain/lang-id-VoxLingua107-ecapa}  & 21    & 11.9          & 6.7         & 7.5  \\
\quad  \textcolor{black}{TitaNet-LID-3x5x512}   
& 12    & 10.7  & 6.3      & 7.0     \\ 
\quad \underline{TitaNet-LID-3x5x1024}                             & 29    & \textbf{7.5}  & 5.2         & 5.6     \\  \midrule
\textit{finetuned from SSL }    &                    &      &      &  \\
\quad wav2vec2.0~\cite{babu2021xls}                             & 300   & 11.5          & 6.3         & 7.2      \\
\quad XLS-R~\cite{babu2021xls}                                  & 300   & 9.1           & \textbf{5.0}& 5.7      \\
\quad XLS-R-attentive~\cite{kukk2022accent}                     & 300   & -             & -           & \textbf{4.7}      \\ \bottomrule
\end{tabular}%
}
\caption{Error rate on VoxLingua107 evaluation set. We compare TitaNet-LID with (1) two other models trained from scratch: Resnet34 and ECAPA-TDNN-LID; (2) three pre-trained self-supervised learning (SSL) models finetuned on VoxLingua107: wav2vec2.0, XLS-R and XLS-R-attentive. Accuracy numbers are computed for different utterance length categories and on all samples.}
\label{table:table_result}
\end{table}

In Table~\ref{table:table_result}, we show that TitaNet-LID-3x5x1024 achieves 5.6\% error rate on all samples, and it achieves 7.5\% and 5.2\% error rate for the short and long recordings respectively. TitaNet-LID gets SOTA performance for models trained from scratch and it has accuracy close to pretrained SSL models while being ten times smaller.
The lowest error rate on 0-5 seconds samples also suggests that TitaNet-LID is robust even on relatively short samples. 
\textcolor{black}{Smaller model TitaNet-LID-3x5x512 with 12M parameters also outperform Resnet34 and ECAPA-TDNN-LID with almost half size comparing to those models.}
\textcolor{black}{TitaNet-LID which employs 1D depth-wise separable convolutions and SE layers could effectively capture local and global context within an utterance which is important for LID task, therefore, it can achieve better result even with less number of parameters.}

\begin{figure}
  \centering
  \centerline{\includegraphics[width=24em]{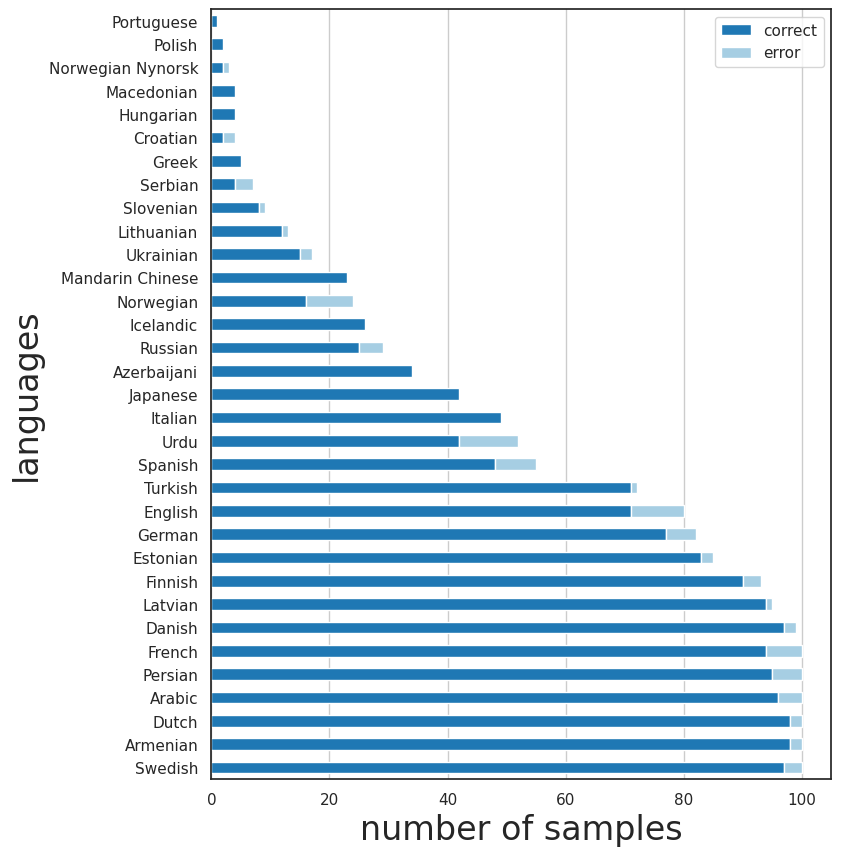}}
  \caption{Number of samples for each language in VoxLingua107 evaluation set. Dark blue represents accurately identified samples of TitaNet-LID-3x5x1024.}
  \label{figure:evaldata}
\end{figure}

In Figure~\ref{figure:evaldata}, we demonstrate the number of the wrongly identified samples of \textcolor{black}{TitaNet-LID-3x5x1024} in each language in the evaluation set with bright blue ink. 
And we can notice that the most wrongly predicted languages are Urdu, English, and Norwegian.
Among the errors, most of them occur between closely related languages such as ``Urdu $\rightarrow$  Hindi'', ``Norwegian $\rightarrow$ Norwegian Nynorsk'' and ``Spanish $\rightarrow$ 
Galician''. 
The English samples that have been wrongly identified as Welsh could possibly be due to the bad quality of the retrieved Welsh data as stated in \cite{valk2021voxlingua107}.

\subsection{Finetune on FLEURS}
\label{ssec:fleurs}

Trained on VoxLingua107, TitaNet-LID has seen 82 out of 102 languages from FLEURS. It has 77.9\% macro accuracy on these seen languages. 
To investigate the ability to adapt TitaNet-LID to unseen languages and acoustic conditions, we conduct two experiments: fine-tune TitaNet-LID-3x5x1024 \textcolor{black}{1) on FLEUR training set, and 2) on the union of training sets} of FLEURS and VoxLingua107.

\subsubsection{Dataset}
\label{sssec:fleurs_data}
FLEURS~\cite{conneau2022FLEURS} was created by recording 2009 sentences from FLoRes-101 benchmark by three different native speakers for each language. 
It contains 102 languages. 
It has official split for LID task where speakers of the train sets are different than speakers from the dev/test sets. 
The sizes of training, dev, and test are 987h, 120h, 283h respectively.
\textcolor{black}{After splitting the utterances into non-overlapping 3-second segments, there are 678K and 83K segments in the FLEURS training and validation sets, respectively. The training sets of FLEURS and VoxLingua107 together encompass 127 languages, while the test sets cover 103 languages. It is worth noting that VoxLingua107 is more diverse and noisy than FLEURS.}

\subsubsection{Fine-tuning setup}
\label{sssec:fleurs_exp}
\textcolor{black}{We freeze TitaNet-LID encoder and change number of output classes in decoder to 102 (FLEURS) and 127 (VoxLingua107 union FLEURS). During fine-tuning, the weight of loss is calculated by occurrence of each class the dataset(s).
Model is fine-tuned for 10 epochs on a node with 2 GPUs. We use the same hyper-parameters as for training step with VoxLingua107, except: a peak learning rate of 5e-5, a dropout of 0.1, and speed perturbation for augmentation. }

\subsubsection{Results}
\label{sssec:fleurs_results}

\begin{table}
\resizebox{\columnwidth}{!}{%
\begin{tabular}{lccccc} \toprule
\multirow{2}{*}{model} & params & \multicolumn{3}{c}{\# languages} & \multirow{2}{*}{\begin{tabular}[c]{@{}c@{}}macro \\ acc \end{tabular}} \\
 & \multirow{2}{*}{(M)}   & pre-        & fine-       & test         &      \\
 &                        & train       & tune        & set         & (\%)
                           \\ \midrule
w2v-bert-51~\cite{conneau2022xtreme}    & 600            & 51              & 102        & 102    & 71.4                                                                            \\
TitaNet-LID  & 29            & 107              & 102      &102       & 88.2                                                                            \\
TitaNet-LID                                & 29             & 107             & 127      &103      & 93.8 \\ \bottomrule    
\end{tabular}%
}
\caption{\textcolor{black}{TitaNet-LID-3x5x1024 finetuned on FLEURS vs pretrained w2v-bert-51 fine-tuned on FLEURS. We also finetune the model on the union of FLEURS and VoxLingua107 (127 languages), achieving higher macro accuracy 93.8\% on 103 languages test set; and the model has the capability to predict 127 languages.}}  
\label{table:FLEURS}
\end{table}

\textcolor{black}{We compare the fintuned models with wav2vec-BERT model, w2v-bert-51 which has 600M parameters and present result in Table~\ref{table:FLEURS}.
w2v-bert-51~\cite{conneau2022xtreme} was first SSL-pretrained on 429k unlabeled data from 51 languages, and then fine-tuned on FLEURS.
It achieves 71.4\% macro accuracy on FLEURS test set. Our TitaNet-LID-3x5x1024 (29M) achieves a 16.8\%  higher accuracy than the 20x larger SSL pre-trained model w2v-bert-51 (0.6B) on the FLEURS test set. Furthermore, we demonstrate that by finetuning our model on the union of FLEURS and VoxLingua107, it can achieve an even higher macro accuracy of 93.8\% on the 103-languages test set, and is capable of predicting 127 languages.}

\begin{table}
  \centering
  \resizebox{\columnwidth}{!}{%
  \begin{tabular}{llll}
    \toprule
    \# samples & \# errors & \multicolumn{2}{c}{error} \\
     &  & True $\rightarrow$ Pred & \# \\
    \midrule
    700 & 570 & Serbian     $\rightarrow$ Bosnian       & 478 \\
    728 & 345 & Javanese    $\rightarrow$ Indonesian    & 344 \\
    299 & 256 & Urdu        $\rightarrow$ Hindi         & 253 \\
    749 & 261 & Malay       $\rightarrow$ Indonesian    & 250 \\
    914 & 337 & Croatian    $\rightarrow$ Serbian       & 229 \\
    998 & 540 & Occitan     $\rightarrow$ Lingala       & 214 \\
    946 & 212 & Asturian    $\rightarrow$ Galician      & 190 \\
    925 & 209 & Bosnian     $\rightarrow$ Croatian      & 189 \\
    927 & 247 & Galician    $\rightarrow$ Catalan       & 170 \\
    980 & 167 & Sindhi      $\rightarrow$ Hindi         & 149 \\
    \bottomrule
  \end{tabular}
  }
  \caption{Most common errors when finetuning on FLEURS only.}
  \label{table:common}
\end{table}
Table~\ref{table:common} shows the most common classification errors observed in the evaluation set in finetuning on FLEURS only experiment. Similar to the observation in VoxLingua107, most of the errors come from related languages such as ``Serbian, Bosnian, and Croatian". The errors such as misclassifying Occitan to be Lingala could possibly be due to the limitation of the model or the inaccuracy of the training and evaluation data. 
\section{Ablation study}
\label{ssec:ablation}

Since LID model is frequently used as a pre-processing step for ASR, it is expected to introduce less latency and extra processing time as possible, especially for online mode. Hence both the input segment length for inference and the model size are crucial parameters.

\subsection{Input length for inference} 
\label{sssec:chunk}
\begin{figure}
  \centering
  \centerline{\includegraphics[width=23em]{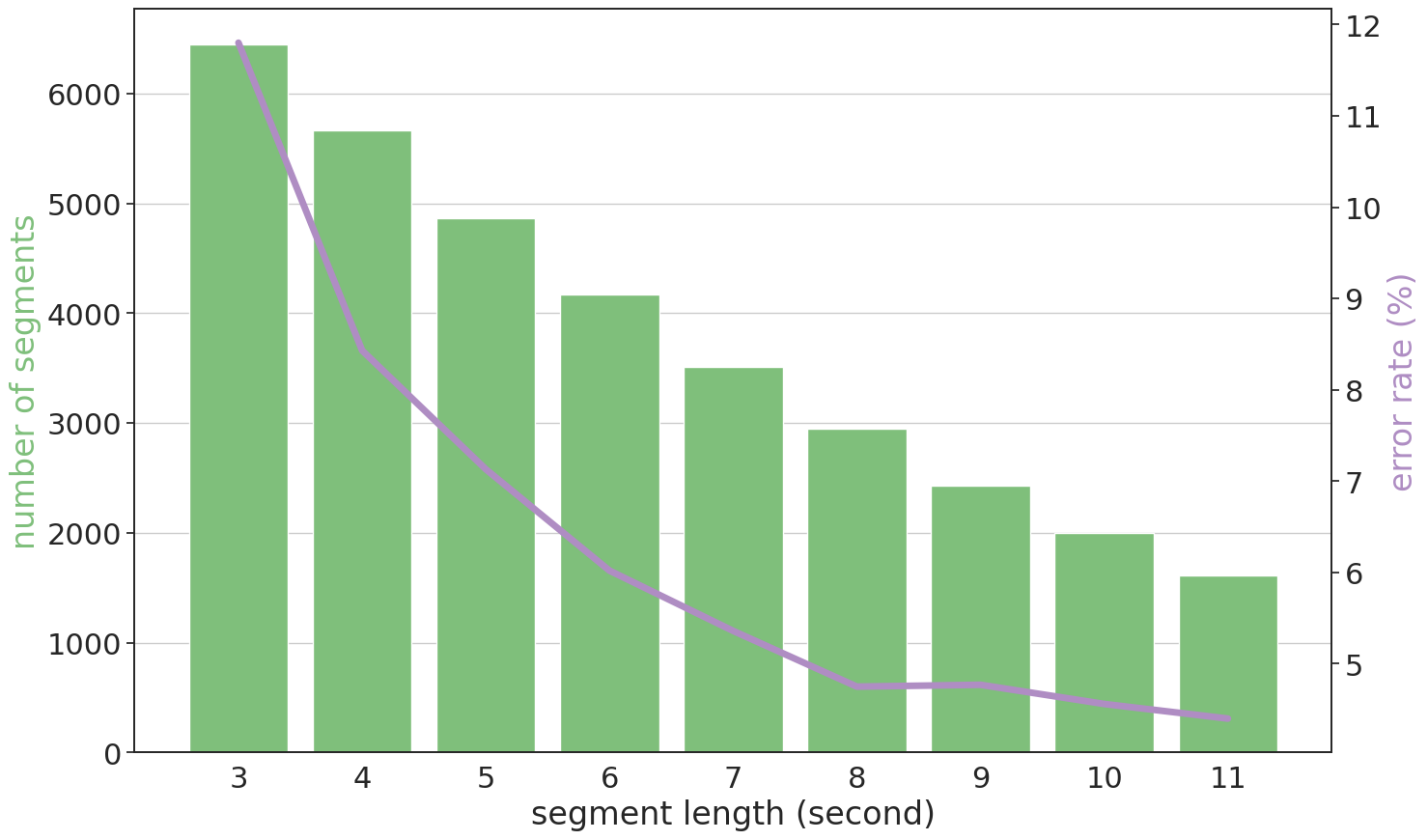}} 
  \caption{The error rate of TitaNet-LID-3x5x1024 on VoxLingua107 evaluation set with different segment length. Generally the larger the input length, the better the model performs.}
  \label{figure:chunk}
\end{figure}

Typically, the longer the segments, the more language information is presented thus leading to higher accuracy of identification.
In this section, we investigate how the input length would impact the performance of the model. 
TitaNet-LID is able to classify the input segments with different lengths by mapping variable-length utterances to a fixed-length representation.
In Figure~\ref{figure:chunk}, we split the evaluation samples into segments of different lengths with stride 2, resulting less number of samples with large segment length. 
It can be seen that even though the model was trained with 3s segments, the model can still perform well and even better with segments of larger lengths. 

\subsection{Model scalability}
\label{ssec:model_size}

\begin{table}
\resizebox{\columnwidth}{!}{%
\begin{tabular}{rrccc} \toprule
\multicolumn{1}{c}{\multirow{2}{*}{R}} & \multicolumn{1}{c}{\multirow{2}{*}{C}} & \multicolumn{1}{c}{Params}  & \multicolumn{1}{c}{val}            & \multicolumn{1}{c}{eval}                  \\
\multicolumn{1}{c}{}                   & \multicolumn{1}{c}{}                   & \multicolumn{1}{c}{(M)}                        & \multicolumn{1}{c}{macro acc (\%)} & \multicolumn{1}{c}{error rate (\%)} \\ \midrule
3  & 256 & 7.3 & 89.8 & 9.6 \\
5  & 256 & 7.7 & 90.7 & 8.4 \\
7  & 256 & 8.1 & 92.1 & 9.2 \\
\midrule
5   & 512  & 12.3   & 93.5 & 7.0 \\
5   & 1024 & 28.9   & 95.0 & 5.6 \\
5   & 2048 & 92.1   & 95.9 & 6.8 \\  \bottomrule     
\end{tabular}%
}
\caption{TitaNet-LID: scalablity study. We use the same number of blocks $B=3$ for all experiments and change model depth using number of sub-blocks $R$ and model width with number of channels $C$ respectively. } 
\label{table:model_size}
\end{table}

The depth and width of TitaNet-LID could be increased or decreased by number of repeated layers R and changing filter size C. 
In this section, we report both validation macro accuracy and evaluation error rate because as shown in Figure~\ref{figure:evaldata}, the evaluation set is quite small and the distribution of it is imbalanced.
As seen in Table~\ref{table:model_size}, the validation macro accuracy would increase alongside the model size. The inconsistent trend of error rate could possibly be due to the limitation of evaluation set.
TitaNet-LID-3x5x512 with 12.3M parameters outperforms the baseline model ECAPA-TDNN-LID with 21M parameters. 
The model could be easily scaled down for faster inference speed for constrained deployment conditions or scaled with for better performance. 

\section{Conclusion}
\label{sec:conclusion}
In this paper, we present TitaNet-LID, an end-to-end model for language identification. It combines 1D depth-wise separable convolutions and Squeeze-and-Excite mechanism with global context to extract intermediary dimensional vector representations. The representations are then sent to a classifier to identify the language of the input audio. 

TitaNet-LID achieves SOTA accuracy on the VoxLingua107 dataset among models trained from scratch. It is close to SOTA for SSL-pretrained models while being 10x fewer parameters. This makes TitaNet-LID an attractive model for memory or compute-constrained environments. 
We also show that TitaNet-LID can be adapted to unseen languages and new acoustic conditions with simple fine-tuning. It achieves new SOTA result on FLEURS.

The model's implementation and checkpoint are made available through NVIDIA NeMo.

\vfill\pagebreak

\bibliographystyle{IEEEtran}
\bibliography{mybib}

\end{document}